\documentclass[english,aps,showpacs,amsmath,amssymb, preprint]{revtex4}
\usepackage[T1]{fontenc}
\usepackage[latin9]{inputenc}
\usepackage{amsthm}
\usepackage{amsbsy}
\usepackage{amstext}
\usepackage{setspace}
\usepackage{amssymb}
\usepackage{graphicx}

\makeatletter
\numberwithin{equation}{section}
\numberwithin{figure}{section}

\makeatother

\makeatother

\usepackage{babel}
\usepackage{graphicx}
\makeatother

\usepackage{babel}

\begin{document}

\title{Asymmetric pulsing for reliable operation of titanium/manganite memristors}

\author{F. Gomez-Marlasca$^{1}$, N. Ghenzi$^{1}$, P. Stoliar$^{1,2,*}$,
M. J. S\'anchez$^{3}$, M. J. Rozenberg$^{4,5}$, G. Leyva$^{1}$, P.
Levy$^{1}$}

\affiliation{$^{1}$GIA, CAC - CNEA, Av. Gral Paz 1499 (1650) San Mart\'{\i}n, Argentina,
\\$^{2}$ECyT, Universidad Nacional de San Mart\'{\i}n, Campus Miguelete,
Mart\'{\i}n de Irigoyen 3100 (1650) San Mart\'{\i}n, Argentina,\\
 $^{3}$Centro At\'omico Bariloche and Instituto Balseiro, CNEA,
(8400) San Carlos de Bariloche, Argentina,\\
 $^{4}$ Laboratoire de Physique des Solides, UMR8502 Universit\'e
Paris-Sud, Orsay 91405, France,\\
 $^{5}$Departamento de F\'{\i}sica Juan Jos\'e Giambiagi, FCEN, Universidad
de Buenos Aires, Ciudad Universitaria Pabell\'on I (1428) CABA, Argentina.}

\begin{abstract}

We present a pulsing protocol that significantly increases the endurance
of a titanium-manganite interface used as a binary memory cell. The
core of this protocol is an algorithm that searches for the proper
values for the set and reset pulses, canceling the drift in the resistance
values. A set of experiments show the drift-free operation for more
than $10^{5}$ switching cycles, as well as the detrimental effect
by changing the amplitude of pulses indicated by the protocol. We
reproduced the results with a numerical model, which provides information
on the dynamics of the oxygen vacancies during the switching cycles. 
\end{abstract}
\maketitle
Metal - transition metal oxide interfaces exhibiting resistive switching
(RS) are prominent candidates for non volatile memory applications.
Despite many appealing features like retentivity, capability of downscaling,
integration in complementary metal-oxide semiconductor (CMOS) 
architectures and speed, one of the main drawbacks
of these systems is the poor endurance.\cite{sawa,Waser2009}

It is nowadays widely accepted that the redistribution of oxygen vacancies
near the metal oxide interface might determine the main features of
the bipolar RS response. \cite{sawa,szot,nian,yang,Jeong2009} A recently
proposed model \cite{rozen}, that succeeded in reproducing non trivial
experimental findings \cite{Ghenzi2010}, indicates that the migration
of oxygen vacancies due to the local electric fields built up in a
nanoscale vicinity of the metal oxide interface is at the origin of
the most significant resistive changes. Each microscopic region of
the sample has a resistivity that is a function of the local oxygen
vacancy concentration. When an electrical pulse (i.e. set) is applied
to the device, local electric fields proportional to the local resistivity
and to the current density develop. If these fields are strong enough,
the oxygen ions will move, changing the vacancies profile along the
interface and hence the total resistance of the device.

After this redistribution of vacancies, a pulse with the opposite
polarity (i.e reset) might not reproduce the initial profile of electric
fields and then, neither the vacancies will return to the original
positions, nor the device will return to the original resistance value
after the completion of the pulsing cycle. The repetition of this
process will produce a drift of the switching resistance levels (ON/OFF
ratio) which eventually fully degrades the memory performance. The
strategies reported in literature to overcome this problem are mainly
based on the design of the device and the selection of materials.\cite{Shen2008,Yang2010,Yang2009,sawa,Shang2010}

In this work we present a pulsing protocol
that allows to obtain more than 10$^{5}$ RS switching cycles in manganite-based
devices, significantly improving their endurance. We compare the results
with numerical simulations.

We have made RS devices of polycrystalline manganite $\mathrm{La_{5/8-y}Pr_{y}Ca_{3/8}MnO_{3}}$
(LPCMO) with Ti contacts\cite{Quintero2004,Levy2002}. The devices,
depicted in the inset of Fig. 1, were fabricated on a LPCMO sintered
bulk 1 mm-thick with a diameter of 13 mm. The 300 nm-thick, 1.5 mm-diameter
pads were sputtered through a shadow mask. The nearest distance between
pads is $\sim$2 mm. The electrical characterization was performed
with one of the Source Measurement Units (SMU) of a Keithley 2602
connected in the remote sense configuration. The SMU applies the writing
(1 ms) and reading (200 ms) pulses through the A and C pads, and measure
the voltage drop between the A and B contacts. In this configuration
the resistance of both A and C metal-oxide interfaces change, but
only the resistance of the A interface is measured.

We define two resistance levels corresponding to two logic levels
required to use the RS devices as a binary memory element, as indicated
in Fig. 1. We define a high level (H) that has to be greater than
R$_{H,min}$ and a low level (L) that always has a resistance lower
than R$_{L,max}$= 0.39$\cdot$R$_{H,min}$. Both values are scaled
depending on the device switching range. The factor 0.39 mimics the
levels of the low-voltage CMOS-logic product family with 
translation (LCX)\cite{LCX}. Defining the levels in
this way introduces a gap in the resistance values. Successful RS
operations are defined as those overcoming this gap.

During the experiments, the generator uses three types of current
pulses: the read pulse that is used to sense the resistance of the
interface, the set pulse that switches the device from the level L
(low resistance) to the level H (high resistance), and the reset pulses
that switches from H to L\cite{nota-notacion inversa}. The read pulse
with  $\mathrm{I_{RD}}$=50 $\mathrm{\mu}$A is low enough to ensure
that no displacement of vacancies occurs (i.e. it does not change
the resistance level).

We implement an algorithm that looks for the proper values (for successful
RS) for the set and reset current pulses, $\mathrm{I_{SET}}$ and
$\mathrm{I_{RST}}$, with the general criteria of keeping a low number
of pulses and the lowest possible amplitude. The algorithm, tries
to set the logic state by applying a single current pulse $\mathrm{I_{SET}^{\prime}}$.
If it fails (i.e. it does not overcome the resistance gap), a second
identical pulse is applied in order to set the state. If this second
attempt fails, the value of $\mathrm{I_{SET}^{\prime}}$ is increased
and the algorithm continues applying pulses with increasing value
until it eventually succeeds. The current is increased in steps. The
values of these steps is fixed through all the experiment and is set
in a few percent of the expected switching current, typically 5-10\%.
The device is considered defective if after increasing the current
value 50 times, the resistance level does not change. If five consecutive
set procedures require these second chances then $\mathrm{I_{SET}^{\prime}}$ will be increased anyway. An analogous criteria is applied for the reset procedure.
Eventually the algorithm finds two independent values for $\mathrm{I_{SET}}$
and $\mathrm{I_{RST}}$ required for stable operation. \cite{supp} 

Fig. 2 shows the initial stage of a typical experiment. Statistically,
the correction to the amplitude of the pulses ceases within the first
3000 cycles, reaching a ratio $|{I_{SET}/I_{RST}}|= 2.6\pm1.2$. Even
if after these first corrections the amplitude of the pulses require
no further corrections, once every 5400$\pm$300 switching attempts
it is necessary to apply a second pulse in order to properly switch
the device. The histogram presented in Fig. 1 depicts the typical
distribution of the resistance values when switching the device with
the proposed protocol during 120000 cycles. Notice that the present
protocol produces no events in the gap.

With the aim of experimentally testing the robustness of the above
described asymmetric pulsing protocol we performed a series of experiments
to show the effect of deliberately changing the ratio $\mathrm{I_{SET}/I_{RST}}$
attained by the protocol (Fig. 3). The devices in these experiments
correspond to different location of the contacts on the same sample
of LPCMO. At the beginning of this set of tests, they were switched
a minimum of 3000 cycles (only the last 300 are shown) in order to
achieve the proper $\mathrm{I_{SET}/I_{RST}}$, depending on the selected
contacts. As a reference, Fig. 3a shows a device that has remained
in operation under the same conditions 700 cycles more. In the device
of Fig. 3b, we stopped the operation of the algorithm and increased
the amplitude of $\mathrm{I_{RST}}$ at the cycle indicated with the
number 0. This reduction of the $\mathrm{I_{SET}/I_{RST}}$ ratio,
results in an evident drift of the resistance values and the H resistance
value level eventually falls inside the gap, i.e. the device no longer
switches between the pre-defined levels. For Fig. 3c the $|\mathrm{I_{SET}/I_{RST}}|$
ratio was increased by a factor 1.5, producing the complementary drift
effect.

We successfully reproduced the experimental results employing the
model introduced in Ref. \cite{rozen}, which is based on the electric
field enhanced oxygen vacancy diffusion dynamics.\cite{supp_mod} In this model the active
region for conduction is a one dimensional resistive network, in which
each site corresponds to a (nanoscale) domain, having a local resistivity
proportional to the local density of oxygen vacancies, $\delta_{i}$.
The dynamic for the vacancies is governed by the equation: \[
p_{i,i+1}=\delta_{i}\left(1-\delta_{i+1}\right)\exp\left(-V_{o}+\Delta V_{i}\right)\;,\]

that specifies the probability for transfer of vacancies between two
neighboring sites. $V_{o}$ is a dimensionless constant related to
the activation energy for vacancies diffusion and $\Delta V_{i}$
is the local potential drop ${\Delta V}_{i}(t)=V_{i+1}(t)-V_{i}(t)$
due to the applied electric pulse. For the present analysis we consider
one interface (I) in contact with the bulk (B) of the sample. The
local resistivity at site $i$ is given by \[
\rho_{i}=\delta_{i}\cdot\left(A_{B}+\frac{A_{I}-A_{B}}{1+\exp\left(\frac{i-N_{I}}{w}\right)}\right)\;,\]

where $A_{B}$ and $A_{I}$ relate vacancy concentration and resistivity
in the bulk and in the interface region respectively ($A_{B}\ll A_{I}$),
and $N_{I}$ and $w$ define the length and the width of the interface.
First, we simulated the application of triangular sweeps to a flat
distribution of vacancies, obtaining hysteresis switching loops similar
to those presented in Ref \cite{Ghenzi2010}. This step emulated the
forming process and provided information on the resistance switching
range of the simulated sample. Then, we simulated the  proposed pulsing protocol. 
We successfully reproduced the continuous cycling
between the L and H states without drift.

The simulation eventually arrived to a stable operation, applying
asymmetrical pulses as in the experiments. For  the actual parameters of
the simulation we obtained a ratio $|\mathrm{I_{SET}/I_{RST}}|=2.9$
that smoothly depends on $w$, but a systematic study of this phenomena
is out of the scope of this work. In stable operation both L and H
states correspond to two well defined profiles of vacancies, $\bar{\boldsymbol{\delta}}_{H}$
and $\bar{\boldsymbol{\delta}}_{L}$. The electrical pulses produce
a redistribution of vacancies 
$\bar{\boldsymbol{\delta}}_{H(L)}{I_{pulse}\atop \longrightarrow}\bar{\boldsymbol{\delta}}_{H(L)}^{\prime}\left(I_{pulse}\right)$,
and a subsequent change in the total resistance of the device $R_{H(L)}\left(\bar{\boldsymbol{\delta}}_{H(L)}\right){I_{pulse}\atop \longrightarrow}R_{H(L)}^{\prime}\left(I_{pulse},\bar{\boldsymbol{\delta}}_{H(L)}^{\prime}\right)$.
Fig. 4 presents the calculated $R_{H(L)}^{\prime}\left(I_{pulse},\bar{\boldsymbol{\delta}}_{H(L)}^{\prime}\right)$,
which was obtained by simulating the application of pulses of different
intensity to $\bar{\boldsymbol{\delta}}_{H}$ and $\bar{\boldsymbol{\delta}}_{L}$. The stable operation is the convergence
of $\mathrm{I_{SET}}$ and $\mathrm{I_{RST}}$ to values where $\bar{\boldsymbol{\delta}}_{L}^{\prime}=\bar{\boldsymbol{\delta}}_{H}$
and $\bar{\boldsymbol{\delta}}_{H}^{\prime}=\bar{\boldsymbol{\delta}}_{L}$.
This condition is presented in Fig. 4: $R_{H}^{\prime}$ equals $R_{L}$
when the amplitude of the pulses is $\mathrm{I_{RST}}$ and $R_{L}^{\prime}=R_{H}$
for pulses of amplitude $\mathrm{I_{SET}}=-2.9\cdot\mathrm{I_{RST}}$.
The lack of symmetry explains the need of non-symmetric pulses to
obtain a stable operation. The drift of the resistance upon symmetric
pulsing is due to the gradual injection of vacancies deeply into the
bulk where they get stuck never returning to the interface region.
The local electric field, $\Delta V_{i}\propto\mathrm{I_{SET(RST)}}$,
is much lower in the bulk region than in the interface. As a consequence,
the amount of vacancies in the interface gradually decreases
with the repetitive cycles, with the concomitant reduction of the
interface resistance. In order that all the vacancies that were injected
into the bulk region during the former set pulse return to the interface
during a subsequent reset pulse, it is necessary to apply a higher
current pulse. If this second current pulse is stronger than required
for a drift-free operation, then the amount of vacancies in the interface
will be slowly increased in each cycle, with the subsequent increment
of the interface resistance.

In conclusion we have proposed an experimental current protocol based
on asymmetric pulsing, that succeeds in finding a stable and repetitive
switching between H an L resistance states up to $10^{5}$ cycles,
improving the endurance of the system by means of virtually canceling
the drift in the resistance values. The theoretical model provides
a physical explanation of the interface resistance degradation in
terms of the injection of vacancies in the bulk region. The proposed
current protocol applied to the model simulations successfully reproduces
the experimental findings, validating the model hypothesis and its
usefulness as a valuable aid in the analysis of the experimental results.

\newpage
\begin{figure}
 \centerline{\includegraphics[width=8cm]{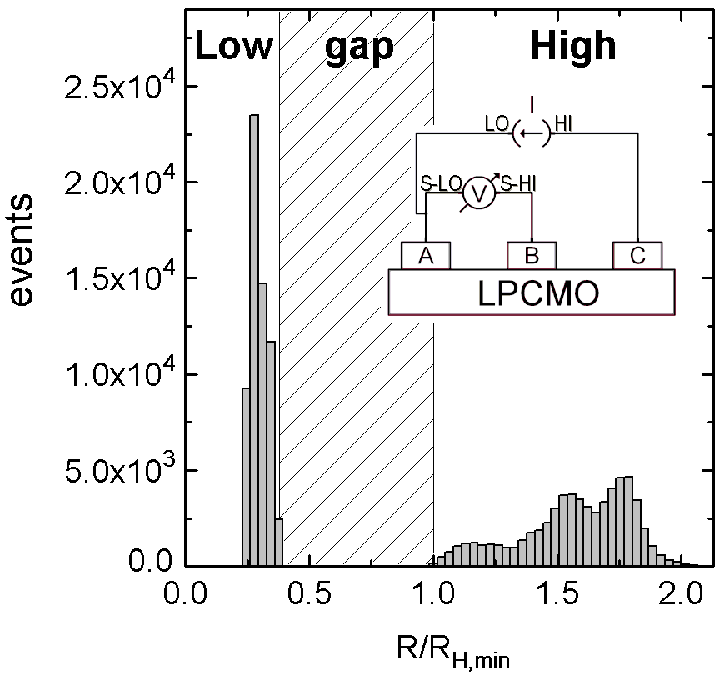}}
\caption{Histogram of the resistance values in a typical experiment
with the definition the H and L levels. The inset shows the experimental
setup.} 
\end{figure}

\begin{figure}
 \centerline{\includegraphics[width=8cm]{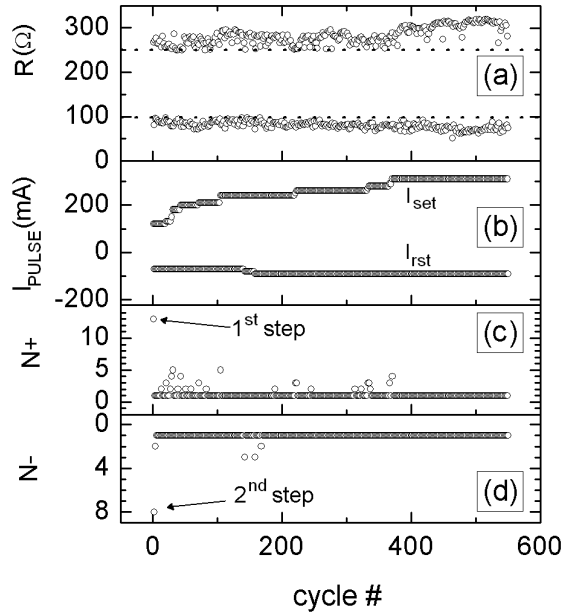}} 
\caption{Initial stage of a typical experiment. The threshold
levels were set to R$_{H,min}$ = 254$\Omega$ and R$_{L,max}$=R$_{H,min}\text{/0.39}$=100$\Omega$,
values compatible with the resistance switching capabilities of the
device (a). The switching protocol starts with the sample in low resistance
level. In the first step, it looks for proper pulse amplitude to set
the device, requiring in this case $N_{+}=14$ positive pulses of
increasing amplitude to overcome the R$_{H,min}$ level (c). The amplitude
of the $14^{th}$ pulse was $+110mA$ (b). In the second step, $N_{-}=8$
negative pulses of increasing amplitude have been needed to reset
the device (d), and so on. After $\sim$400 cycles the correction
to the pulsing amplitudes ceases, reaching a ratio ${I_{SET}/I_{RST}}=310mA/-90mA$.} 
\end{figure}

\begin{figure}
 \centerline{\includegraphics[width=8cm]{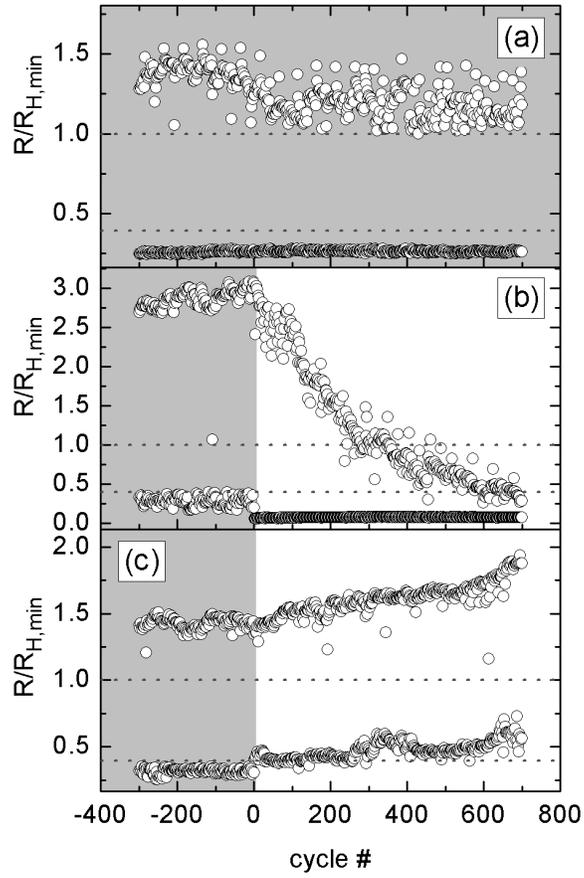}} 
\caption{Effect of changing the $\mathrm{I_{SET}/I_{RST}}$
ratio when the device is in stable operation. Shaded in gray, the
region were program operates. The horizontal dotted lines indicate
R$_{L,max}$ and R$_{H,min}$ values. a) figure for reference, no
change in the $\mathrm{I_{SET}/I_{RST}}$ ratio. b) the initial $\mathrm{I_{SET}/I_{RST}}=110/-20$
changed to 110/-90 at cycle \#0. c) $\mathrm{I_{SET}/I_{RST}}$ changed
from 310/-90 to 310/-60.} 
\end{figure}

\begin{figure}
 \centerline{\includegraphics[width=8cm]{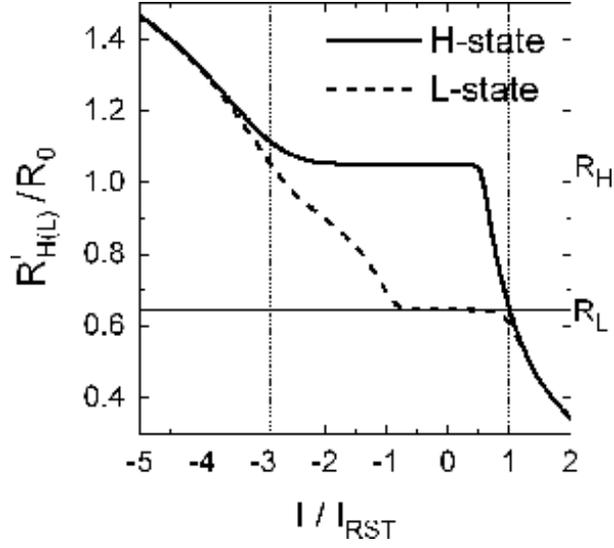}} 
\caption{Model simulation of the achieved resistance
for stable operation. The pulse amplitude required to change the vacancy distribution of the
L state to the vacancy distribution of the H state is $I_{SET}=-2.9\cdot I_{RST}$.
The resistance values are normalized to the resistance of a sample
with the same number of vacancies uniformly distributed, $R_{0}$.} 
\end{figure}

\begin{thebibliography}{16}
\bibitem{nota-notacion inversa} Notice this is not the standard definition
of low and high levels in RS.
\end{thebibliography}
\end{document}